\documentclass[prd,twocolumn,twoside,preprintnumbers,superscriptaddress,nofootinbib]{revtex4}
\usepackage{graphicx}
\usepackage{multirow}

\usepackage{siunitx}
\usepackage{tikz}

\def\beq{\begin{equation}}
\def\eeq{\end{equation}}
\def\bea{\begin{eqnarray}}
\def\eea{\end{eqnarray}}
\def\nn{\nonumber}
\def\roughly#1{\mathrel{\raise.3ex\hbox
{$#1$\kern-.75em\lower1ex\hbox{$\sim$}}}}

\def\bd{B^0}
\def\bs{B_s^0}

\def\pip{\pi^+}
\def\pim{\pi^-}

\def\order{\lower 1.8ex \hbox{\LARGE\~{}}}

\def \<{\left<}
\def \>{\right>}
\def \[{\left[}
\def \]{\right]}
\def \({\left(}
\def \){\right)}

\def \l.{\left.}
\def \r.{\right.}

\def\text#1{{\rm #1}}

\def\kp{K^+}
\def\km{K^-}
\def\norm#1{\left|#1\right|}

\def\me#1#2#3{\left\langle#1\left|\left|#2\right|\right|#3\right\rangle}
\def\lam#1#2#3{\lambda^#1_{#2#3}}
\def\Ta#1{T_#1^1}
\def\Pa#1{P_#1^1}
\def\Tb#1{T_#1^0}
\def\Pb#1{P_#1^0}

\begin{document}

\preprint{UdeM-GPP-TH-22-297}

\title{\boldmath A U-Spin Puzzle in $B$ Decays}

\author{Bhubanjyoti Bhattacharya}\email{bbhattach@ltu.edu}
\affiliation{Department of Natural Sciences, Lawrence Technological University, Southfield, MI 48075, USA}

\author{Suman Kumbhakar}\email{suman.kumbhakar@umontreal.ca}
\affiliation{Physique des Particules, Universit\'e de Montr\'eal,
  Montr\'eal, QC, Canada  H2V 0B3}

\author{David London}\email{london@lps.umontreal.ca}
\affiliation{Physique des Particules, Universit\'e de Montr\'eal,
  Montr\'eal, QC, Canada  H2V 0B3}

\author{Nicolas Payot}\email{nicolas.payot@umontreal.ca}
\affiliation{Physique des Particules, Universit\'e de Montr\'eal,
  Montr\'eal, QC, Canada  H2V 0B3}

\date{\today}

\begin{abstract}
We impose U spin symmetry ($SU(2)_{\rm Uspin}$) on the Hamiltonian for $B$ decays. As expected, we find the equality of amplitudes related by the exchange $d \leftrightarrow s$. We also find that the amplitudes for the $\Delta S=0$ processes $\bd \to \pi^+\pi^-$, $\bs\to\pip\km$ and $\bd\to\kp\km$ form a U-spin triangle relation. The amplitudes for $\bs\to\kp\km$, $\bd\to\pim\kp$ and $\bs\to\pip\pim$ form a similar $\Delta S=1$ triangle relation. And these two triangles are related to one another by $d \leftrightarrow s$. We perform fits to the observables for these six decays. If perfect U spin is assumed, the fit is very poor. If U-spin-breaking contributions are added, we find many scenarios that can explain the data. However, in all cases, 100\% U-spin breaking is required, considerably larger than the naive expectation of $\sim 20\%$. This is the U-spin puzzle; it may be strongly hinting at the presence of new physics.
\end{abstract}

\maketitle 


In Ref.~\cite{Fleischer:1999pa}, a method was proposed to extract the angle $\gamma$ of the unitarity triangle from measurements of $\bd \to \pi^+\pi^-$ and $\bs \to K^+ K^-$, two decays related by the exchange $d \leftrightarrow s$. This is usually referred to as U spin symmetry.

Now, U spin is more than just a $d \leftrightarrow s$ symmetry.
It is based on the group $SU(2)_{\rm Uspin}$. Here we examine the
consequences of imposing this symmetry group on the Hamiltonian for $B$ decays. We
find that the $d \leftrightarrow s$ symmetry is reproduced as
expected, but other useful relations also appear. In particular, the
amplitudes for the $\Delta S=0$ processes $\bd \to \pi^+\pi^-$,
$\bs\to\pip\km$ and $\bd\to\kp\km$ form a U-spin triangle relation
similar to the isospin $B \to\pi\pi$ triangle relation \cite{Gronau:1990ka}. Similarly, the amplitudes for
$\bs\to\kp\km$, $\bd\to\pim\kp$ and $\bs\to\pip\pim$ form a $\Delta
S=1$ U-spin triangle relation. And these two triangles are related to
one another by the $d \leftrightarrow s$ symmetry. All six decays have
been measured. In this paper, we show that a simultaneous analysis of their
observables has some puzzling results.

Under $SU(2)_{\rm Uspin}$, $(d,s)$ is a doublet and $({\bar s}, -{\bar d})$ is its conjugate. The mesons that are eigenstates of U spin are then
\bea
& K^0 = d {\bar s} ~~,~ U^0 = \frac{1}{\sqrt{2}} (s {\bar s} - d {\bar d}) ~~,~ {\bar K}^0 = - s {\bar d} ~, & \nn\\
& K^+ = u {\bar s} ~~,~ \pi^+ = -u {\bar d} ~~,~ \pi^- = d {\bar u} ~~,~ K^- = s {\bar u} ~, & \nn\\
& U_8 = \frac{1}{\sqrt{6}} (2 u {\bar u} - d {\bar d} - s {\bar s}) ~. &
\label{Uspinmesons}
\eea
Thus, under $SU(2)_{\rm Uspin}$, $(K^0, U^0 , {\bar K}^0)$ form a
triplet, $(\pi^-, K^-)$ and $(K^+, \pi^+)$ are doublets, and $U_8$ is
a singlet. 

Consider $B \to PP$ decays in the U-spin basis ($P$ is a pseudoscalar meson). Here the initial state is either a doublet [$(\bd,\bs)$] or a singlet [$B^+$], and the final state is one of TT, DD, TD, TS, DS, or SS, where T, D and S refer to a U-spin triplet, doublet and singlet, respectively. The weak Hamiltonian involves $b \to q u {\bar u}$ and $b \to q$ ($q = d$ or $s$), which has $U=\frac12$ for both $q=d$ and $q=s$. However, note that $\Delta S = 0$ decays ($q=d$) and $\Delta S = 1$ decays ($q=s$) involve different Cabibbo-Kobayashi-Maskawa (CKM) matrix elements. 

It is now straightforward to compute the $B \to PP$ decay amplitudes in terms of the $SU(2)_{\rm Uspin}$ reduced matrix elements (RMEs). As the $U^0$ and $U_8$ are both linear combinations of the $\pi^0$, $\eta$ and $\eta'$ mesons, but with unknown relative strong phases, decays involving these particles are not very useful. It is more interesting to consider instead decays whose final states involve only $\pi^\pm$, $K^\pm$, $K^0$ and ${\bar K}^0$. 

In this paper, we focus specifically on $B \to {\rm DD}$ decays. There are three decays with $\Delta S=0$ and three with $\Delta S=1$. In terms of U-spin RMEs, the amplitudes are
\bea
\label{BDDamps_DeltaS0}
\Delta S = 0 ~:
&& A(\bd\to\pip\pim) = M^{\frac12}_{1d} + M^{\frac12}_{0d} ~, \nn\\
&& A(\bd\to\kp\km) = M^{\frac12}_{1d} - M^{\frac12}_{0d} ~, \nn\\
&& A(\bs\to\pip\km) = 2 M^{\frac12}_{1d} ~, \\
\label{BDDamps_DeltaS1}
\Delta S = 1 ~:
&& A(\bs\to\kp\km) = M^{\frac12}_{1s} + M^{\frac12}_{0s} ~, \nn\\
&& A(\bs\to\pip\pim) = M^{\frac12}_{1s} - M^{\frac12}_{0s} ~, \nn\\
&& A(\bd\to\pim\kp) = 2M^{\frac12}_{1s} ~,
\eea
where the subscripts $d$ and $s$ refer respectively to the transitions $b \to
d$ ($\Delta S = 0$) and $b \to s$ ($\Delta S = 1$), and
$M^{\frac12}_{1q} \equiv \me{1}{H^{\frac12}}{\frac12}_q$, $M^{\frac12}_{0q} \equiv \me{0}{H^{\frac12}}{\frac12}_q$, $q=d,s$. (Note that we have absorbed the magnitudes of the Clebsch-Gordan coefficients into the RMEs.)

The amplitudes $A(\bd\to\pip\pim)$ and $A(\bs\to\kp\km)$ are related to one another by the exchange
$d\leftrightarrow s$, as are the pairs $(A(\bd\to\kp\km), A(\bs\to\pip\pim))$ and
$(A(\bs\to\pip\km), A(\bd\to\pim\kp))$. But there is more: under
$SU(2)_{\rm Uspin}$ symmetry, there are two triangle relations, each
involving only $\Delta S=0$ or $\Delta S=1$ amplitudes\footnote{Similar relations have been derived in Ref.~\cite{Gronau:1994rj}, but in a different context.}:
\bea
A(\bd\to\pip\pim) + A(\bd\to\kp\km) && \nn\\
&& \hskip-3truecm =~A(\bs\to\pip\km) ~, \nn\\
A(\bs\to\kp\km) + A(\bs\to\pip\pim) && \nn\\
&& \hskip-3truecm =~A(\bd\to\pim\kp) ~.
\label{Uspintrianglerelations}
\eea
As we will see below, these can be used to extract a
great deal of information, including the CP phase $\gamma$.

The $B \to {\rm DD}$ amplitudes are given in terms of U-spin RMEs in
Eqs.~(\ref{BDDamps_DeltaS0}) and (\ref{BDDamps_DeltaS1}). Now,
\bea
M^{\frac12}_{0d}
&=& V^*_{ub}V_{ud}\me{0}{H^{\frac12}}{\frac12}^u \nn\\
&& \hskip-12truemm +~V^*_{cb}V_{cd}\me{0}{H^{\frac12}}{\frac12}^c + V^*_{tb}V_{td}\me{0}{H^{\frac12}}{\frac12}^t \nn\\
& \equiv & \lam{u}{b}{d} \, T^0_d + \lam{c}{b}{d} \, P^0_d ~,
\label{bdRMEdivide}
\eea
where $\lam{i}{b}{j} = V^*_{ib}V_{ij}$ ($i=u,c$, $j=d,s$), and we have
used the unitarity of the CKM matrix in passing from the first line to
the second. (Note that, despite the notation, $T^0_d$ and $P^0_d$ do not necessarily correspond only to tree and penguin contributions, respectively.) Similarly,
\bea
& M^{\frac12}_{1d} = \lam{u}{b}{d} \, T^1_d + \lam{c}{b}{d} \, P^1_d ~, & \\
& M^{\frac12}_{0s} = \lam{u}{b}{s} \, T^0_s + \lam{c}{b}{s} \, P^0_s ~,~~ 
M^{\frac12}_{1s} = \lam{u}{b}{s} \, T^1_s + \lam{c}{b}{s} \, P^1_s ~. & \nn
\label{bsRMEdivide}
\eea
With this, the six $B \to {\rm DD}$ decay amplitudes are given by
\newpage
\bea
\label{DeltaS=0amps}
&& \Delta S = 0 ~: \nn\\
&& \hskip-5truemm A(\bd\to\pip\pim) = 
\lam{u}{b}{d}\Ta{d} + \lam{c}{b}{d}\Pa{d} + \lam{u}{b}{d}\Tb{d} + \lam{c}{b}{d}\Pb{d} ~, \nn\\
&& \hskip-5truemm A(\bd\to\kp\km) =
\lam{u}{b}{d}\Ta{d} + \lam{c}{b}{d}\Pa{d} - \lam{u}{b}{d}\Tb{d} - \lam{c}{b}{d}\Pb{d} ~, \nn\\
&& \hskip-5truemm A(\bs\to\pip\km) = 2\lam{u}{b}{d}\Ta{d} + 2\lam{c}{b}{d}\Pa{d} ~, \\
\label{DeltaS=1amps}
&& \Delta S = 1 ~: \nn\\
&& \hskip-5truemm A(\bs\to\kp\km) = \lam{u}{b}{s}\Ta{s} + \lam{c}{b}{s}\Pa{s}
+ \lam{u}{b}{s}\Tb{s} + \lam{c}{b}{s}\Pb{s} ~, \nn\\
&& \hskip-5truemm A(\bs\to\pip\pim) = \lam{u}{b}{s}\Ta{s} + \lam{c}{b}{s}\Pa{s}
- \lam{u}{b}{s}\Tb{s} - \lam{c}{b}{s}\Pb{s} ~, \nn\\
&& \hskip-5truemm A(\bd\to\pim\kp) = 2\lam{u}{b}{s}\Ta{s} + 2\lam{c}{b}{s}\Pa{s} ~.
\eea

These decays have all been measured, yielding a number of observables.
The results of the present experimental measurements
are shown in Table \ref{BtoDD:expValues}. For each of the four decays
$\bd\to\pip\pim$, $\bs\to\pip\km$, $\bs\to\kp\km$ and $\bd\to\pim\kp$,
the branching ratio and direct and indirect (where applicable) CP
asymmetries have been measured. For the rarer decays $\bd\to\kp\km$ and $\bs\to\pip\pim$, we have only the branching ratios. 

\begin{table}[h!]
    \centering
    \begin{tabular}{|c|c|}
    \hline
     Decay    & Observable  \\
     \hline
\multicolumn{2}{|c|}{$\Delta S=0$} \\
\hline
  $\bd\to\pi^+\pi^-$ & $\mathcal{B} = (5.15 \pm 0.19) \times 10^{-6}$ \cite{Belle:2012dmz, LHCb:2012ihl}
 \\
  & $A^{CP} = 0.311 \pm 0.030$ \cite{LHCb:2020byh} \\
  & $S^{CP} = -0.666 \pm 0.029$ \cite{LHCb:2020byh} \\
\hline
  $\bd\to K^+K^-$ & $\mathcal{B} = (8.0 \pm 1.5) \times 10^{-8}$ \cite{LHCb:2016inp} \\
\hline
  $\bs\to\pi^+K^-$ & $\mathcal{B} = (5.9 \pm 0.9) \times 10^{-6}$ \cite{CDF:2008llm, LHCb:2012ihl} \\
  & $A^{CP} = 0.225 \pm 0.012$ \cite{LHCb:2020byh} \\
\hline
\multicolumn{2}{|c|}{$\Delta S=1$} \\
\hline
  $\bs\to K^+K^-$ & $\mathcal{B} = (2.66 \pm 0.32) \times 10^{-5}$ \cite{LHCb:2016inp,CDF:2011ubb} \\
        & $A^{CP} = -0.17 \pm 0.03$  \cite{LHCb:2020byh} \\
        & $S^{CP} = 0.14 \pm 0.03$ \cite{LHCb:2020byh} \\
\hline
  $\bs\to\pi^+\pi^-$ & $\mathcal{B} = (7.2 \pm 1.1) \times 10^{-7}$ \cite{LHCb:2016inp} \\
\hline
   $\bd\to\pi^-K^+$ & $\mathcal{B} = (1.95 \pm 0.05) \times 10^{-5}$ \cite{Belle:2012dmz, BaBar:2006pvm}\\
   & $A^{CP} = -0.0836 \pm 0.0032$ \cite{LHCb:2020byh} \\      
     \hline
    \end{tabular}
    \caption{Experimental values of $B \to$ DD observables. Here, $\mathcal{B}$, $A^{CP}$ and $S^{CP}$ refer to the branching ratio, direct CP asymmetry and indirect CP asymmetry, respectively. The average values given above are taken from Ref.~\cite{HFLAV:2022pwe}. These average values are generally dominated by a few measurements, whose references are given in the Table.}
 \label{BtoDD:expValues}
 \end{table}

The three $\Delta S = 0$ amplitudes [Eq.~(\ref{DeltaS=0amps})] and three $\Delta S = 1$ amplitudes [Eq.~(\ref{DeltaS=1amps})] each involve seven unknown hadronic parameters: the four magnitudes $|\Ta{q}|$, $|\Pa{q}|$, $|\Tb{q}|$ and $|\Pb{q}|$ ($q=d,s$),
along with three relative strong phases. (We take the magnitudes of the CKM matrix elements
from independent measurements \cite{Workman:2022ynf}.) The weak phase $\gamma$ in
$\lam{u}{b}{q}$ is present in all amplitudes; it can be allowed to vary or be constrained by its independently-measured value of $(65.9^{+3.3}_{-3.5})^{\circ}$ \cite{HFLAV:2022pwe,Workman:2022ynf}.
In the U-spin limit, the $\Delta S = 0$ and $\Delta S = 1$ hadronic parameters are equal, so that
all six amplitudes involve the same eight theoretical
parameters. With twelve observables in Table \ref{BtoDD:expValues}, one can perform a fit to the data. In this way, we can determine
the preferred sizes of these parameters, allowing us to extract
$\gamma$ and/or ascertain how well the hypothesis of U-spin symmetry
holds up.

In this fit, we allow $\gamma$ to be a free parameter and, without loss of generality, we take $\delta_{T^1_d} = 0$. The fit is performed using the program {\tt MINUIT} \cite{James:1975dr,James:2004xla,James:1994vla}; the results are shown in Table \ref{6ampfit, gamma free}. Although the best-fit value of $\gamma$ is very close to its present value, this is unimportant, as the fit is very poor: it has $\chi^2_{\rm min}/{\rm d.o.f.} = 17.8/4$, for a p-value of 0.001. 

\begin{table}[h!]
    \centering
    \begin{tabular}{|c|c|}
    \hline
     Parameter    & Best fit Value  \\
     \hline
    $|T^1_d|$  &  $3.85  \pm  0.22$\\
    \hline
    $|P^1_d|$ & $0.56  \pm  0.02$\\
    \hline
    $|T^0_d|$ & $3.27   \pm  0.24$\\
    \hline
    $|P^0_d|$ & $0.71  \pm   0.13$\\
    \hline
    $\delta_{P^1_d}$ & $0.33 \pm  0.01$\\
    \hline
    $\delta_{T^0_d}$ & $0.14  \pm  0.09 $\\
    \hline
    $\delta_{P^0_d}$ & $0.59 \pm  0.20$\\
    \hline
    $\gamma$ & $(67.6 \pm 3.4)^{\circ}$\\
    \hline	
    \end{tabular}
    \caption{Results of a fit to the observables of the six $B \to
      {\rm DD}$ decays in the U-spin limit.  Amplitudes are given in
      keV and phases (apart from $\gamma$) are given in radians.}
\label{6ampfit, gamma free}
\end{table}

It is instructive to search for the reason(s) for this poor fit. We find that there are two ingredients. The first is the ``U-spin relation.'' In the U-spin limit, the observables associated with pairs of decays related by $d \leftrightarrow s$ obey the following relation \cite{Gronau:2013mda}:
\bea
-\frac{A_s^{CP}}{A_d^{CP}} \, \frac{\tau(B_d) \, \mathcal{B}_s}{\tau(B_s) \, \mathcal{B}_d} = 1 ~.
\label{Uspinrelation}
\eea
Here, $B_d$ is the decaying $B$ meson in the $\Delta S = 0$ process,
$\tau(B_d)$ is its lifetime, $A_d^{CP}$ is the direct CP asymmetry in
the decay, and $\mathcal{B}_d$ is the branching ratio. The analogous
quantities for the $\Delta S = 1$ process are indicated by the
subscript $s$. The extent to which the U-spin relation is violated gives a handle on the
size of U-spin breaking.

The second ingredient is more subtle. From Table \ref{BtoDD:expValues}, we see that the branching ratio of $\bd\to K^+K^-$ is much smaller than those of the other $\Delta S=0$ decays. Similarly, 
$\bs\to\pi^+\pi^-$ has by far the smallest branching ratio of the $\Delta S=1$ decays. 

Consider the limit in which these branching ratios are set to zero. 
This approximation is equivalent to setting the $\bd\to K^+K^-$
and $\bs\to\pi^+\pi^-$ amplitudes to zero. (In the diagrammatic language of Ref.~\cite{Gronau:1994rj}, this corresponds to neglecting the subdominant diagrams $E$ and $PA$.) At the RME level, this
implies $\Ta{q} = \Tb{q}$ and $\Pa{q} = \Pb{q}$, $q=d,s$ [see
  Eqs.~(\ref{DeltaS=0amps},\ref{DeltaS=1amps})]. In this limit, we
have
\bea
&& \Delta S = 0 ~:~ A(\bd\to\pip\pim) = A(\bs\to\pip\km) \nn\\
&& \hskip2truecm =~2\lam{u}{b}{d}\Ta{d} + 2\lam{c}{b}{d}\Pa{d} ~, \nn\\
&& \Delta S = 1 ~:~ A(\bs\to\kp\km) = A(\bd\to\pim\kp) \nn\\
&& \hskip2truecm =~2\lam{u}{b}{s}\Ta{s} + 2\lam{c}{b}{s}\Pa{s} ~.
\label{neglectBRs}
\eea

First, this implies that each of the $\Delta S = 0$ amplitudes
is related to each of the $\Delta S = 1$ amplitudes by the exchange
$d \leftrightarrow s$. That is, the U-spin relation
[Eq.~(\ref{Uspinrelation})] applies to four pairs of decays. For all
four pairs, in Table \ref{tab:USpinRelation} we present the values of
the U-spin relation obtained from the experimental data.
  
\begin{table}[!h]
        \centering
        \setlength\extrarowheight{3pt}
        \begin{tabular}{|c|c|c|}
            \hline
            $\Delta S=0$ Decay & $\Delta S = 1$ Decay & U-spin Relation \\
            \hline
            $\bd\to\pi^+\pi^-$ & $\bs\to K^+K^-$ & $2.78\pm0.66$ \\
            $\bd\to\pi^+\pi^-$ & $\bd\to\pi^-K^+$ & $1.02\pm0.12$ ~~ (*) \\
            $\bs\to\pi^+K^-$ & $\bs\to K^+K^-$ & $3.41\pm0.91$ ~~ (*) \\
            $\bs\to\pi^+K^-$ & $\bd\to\pi^-K^+$ & $1.25\pm0.21$ \\
            \hline
        \end{tabular}
        \caption{Values of the U-spin relation for different pairs of
          decays. Entries marked with $(*)$ correspond to pairs
          related only when the small branching ratios are set to zero [Eq.~(\ref{neglectBRs})].}
        \label{tab:USpinRelation}
    \end{table}

These values are to be compared with the ``prediction'' of 1 for this quantity. (For two pairs, the prediction is approximate, as it results from setting the small branching ratios to zero.) We see that the two entries involving $\bs\to K^+K^-$ are in disagreement with the prediction. On the other hand, the two entries with $\bd\to\pi^-K^+$ are in good agreement.

Second, the triangle relations of Eq.~(\ref{Uspintrianglerelations}) become simple
amplitude equalities: 
\bea
A(\bd\to\pip\pim) &=& A(\bs\to\pip\km) ~, \nn\\
A(\bs\to\kp\km) &=& A(\bd\to\pim\kp) ~. 
\eea
These equalities apply to the CP-conjugate amplitudes as well. Using
the measured values of the branching ratios and direct CP asymmetries
for the decays (Table \ref{BtoDD:expValues}), one can extract the magnitudes of $A$ and ${\bar A}$
for each decay. We find
\bea
& A_1 = A(\bs\to\pip\km) ~,~~ A_2 = A(\bd\to\pip\pim) & \nn\\
& \left\vert \frac{\displaystyle A_1}{\displaystyle A_2} \right\vert = 1.05\pm0.08 ~,~~ 
\left\vert \frac{\displaystyle \bar{A}_1}{\displaystyle \bar{A}_2} \right\vert = 1.15\pm0.09 ~, & \nn\\
& A_3 = A(\bd\to\pim\kp) ~,~~ A_4 = A(\bs\to\kp\km) & \nn\\
& \norm{\frac{\displaystyle A_3}{\displaystyle A_4}} = 0.89\pm0.06 ~,~~ 
\norm{\frac{\displaystyle \bar{A}_3}{\displaystyle \bar{A}_4}} = 0.81\pm0.05 ~. & 
\eea
Given the error implicit in the ``prediction'' of 1 for these quantities, these results show no obvious disagreements. Still, it is interesting to note that, once again, it is the ratios involving
$\bs\to K^+K^-$ that exhibit the largest differences from 1.

We have therefore identified certain tensions in the data that may contribute to the poor fit of Table \ref{6ampfit, gamma free}. The only possible way to improve the fit is to include U-spin-breaking contributions. But this may be somewhat delicate: since not all relations predicted by U spin (and/or the neglect of the small branching ratios) are broken, adding U-spin-breaking effects to correct one problem may create another problem where none existed before.

In addition to the seven hadronic parameters of Table \ref{6ampfit, gamma free}, there are nine U-spin-breaking parameters. We define\footnote{Note that U-spin breaking can also be defined at the level of observables, see, for example, Ref.~\cite{Grossman:2013lya}.}
\bea
&& T^0_s = T^0_d (1 + t_0 e^{i\delta_{t_0}}) ~,~~ T^1_s = T^1_d (1 + t_1 e^{i\delta_{t_1}}) ~, \nn\\
&& P^0_s = P^0_d (1 + p_0 e^{i\delta_{p_0}}) ~,~~ P^1_s = P^1_d (1 + p_1 e^{i\delta_{p_1}}) ~, \nn\\
&& A(\bd\to\pip\pim) + A(\bd\to\kp\km) \nn\\
&& \hskip2.5truecm =~ (1+X) A(\bs\to\pip\km) ~, \nn\\
&& A(\bs\to\kp\km) + A(\bs\to\pip\pim) \\
&& \hskip2.5truecm =~ (1+X)A(\bd\to\pim\kp) ~. \nn
\eea
In the fits, we generally constrain the phase $\gamma$ by its measured value. However, this could be incorrect in the presence of new physics. In light of this possibility, we occasionally allow $\gamma$ in $\Delta S = 1$ decays ($\gamma_1$) to be a free parameter.

With twelve observables, there is room for five additional unknown parameters in the fit. Since there are an infinite number of possibilities for these five (linear combinations of) parameters, we cannot draw a definitive conclusion. However, we have examined many sets of five parameters, and certain patterns have emerged.

With the addition of these parameters, we have twelve equations in twelve unknowns. We search for a solution by doing a fit. If $\chi^2_{\rm min} = 0$ is found, this corresponds to an exact solution. We make the following observations:
\begin{itemize}

\item If $t_0$ is not included, we find no solutions.

\item If $t_0$ is included, but is real (i.e., $\delta_{t_0} = 0$), we find no solutions.

\item If $t_0$ and $\delta_{t_0}$ are included, but $t_0$ is combined with another parameter (e.g., $p_0 = t_0$, $p_1 = t_0$ or $t_1 = t_0$), we find no solutions.

\item We find a number of solutions with $t_0$ and $\delta_{t_0}$ nonzero; in all of them, the other magnitudes of U-spin-breaking parameters ($t_1$, $p_0$, $p_1$, $X$) are small.

\item It is not necessary that $\gamma_1$ be included in order to find a solution. However, if it is included, there are solutions, and in all cases $\gamma_1$ is different from its measured value.

\end{itemize}

These properties can be seen even in fits with fewer than five U-spin-breaking parameters, see Table \ref{6ampfit, Uspinbreak}. When only $t_0$ is included, the fit is poor. But it becomes passable when $\delta_{t_0}$ is added, and good with one more parameter. A good fit can be found if $\gamma_1$ is included and allowed to vary, though this is not absolutely necessary. When $\gamma_1$ is included, its best-fit value is found to be different from the measured value of $\gamma$.

\begin{table}[!h]
        \centering
        \setlength\extrarowheight{3pt}
        \begin{tabular}{|c|c|c|}
            \hline
            Parameter & $\chi^2_{\rm min}/{\rm d.o.f.}$ & p-value \\
            \hline
            $t_0 = 0.5 \pm 0.4$ & $16.4/4$ & $0.003$ \\
            \hline
            $t_0 = 1.25 \pm  0.35 $ & & \\
            $\delta_{t_0} = -1.27 \pm  0.33$ & $6.1/3$ & $0.11$ \\
            \hline
            $t_0 = 1.15 \pm   0.34$ & & \\
            $\delta_{t_0} = -1.22  \pm  0.34$ & $1.1/2$ & $0.58$ \\
            $p_1 = 0.28 \pm 0.15$ & & \\
             \hline
            $t_0 = 1.02 \pm 0.31$ & & \\
            $\delta_{t_0} = -1.5  \pm  0.4$ & $1.7/2$ & $0.43$ \\
            $\gamma_1 = (91.1 \pm 14.9)^\circ$ & & \\
              \hline
        \end{tabular}
  \caption{Results of fits to the observables of the six $B \to
      {\rm DD}$ decays in which some U-spin-breaking parameters have been included. Amplitudes are given in keV and phases are given in radians ($\delta_{t_0}$) or degrees ($\gamma_1$).}  
  \label{6ampfit, Uspinbreak}
    \end{table}

But the key point is that, in all the fits that account for the data reasonably well, $t_0 = O(1)$. That is, 100\% U-spin breaking is required, specifically in the $T^0_q$ RME. This U-spin breaking is considerably larger than the naive expectation of $f_K/f_\pi - 1 = ~\sim 20\%$. This is the U-spin puzzle in $B$ decays. (Interestingly, large U-spin breaking has also been observed in $D$ decays, see Refs.~\cite{Schacht:2022kuj, Bause:2022jes}.)

What can be the explanation for this U-spin breaking? There are no known mechanisms in the standard model that can generate U-spin breaking this large; this may be strongly hinting at new physics. New-physics contributions to U-spin breaking have been explored in Ref.~\cite{Nir:2022bbh}. 

At present, there are other hints of new physics in $b \to s$ transitions: in certain observables involving the transition $b \to s \mu^+\mu^-$ \cite{London:2021lfn} and in $B \to \pi K$ decays \cite{Beaudry:2017gtw}. The result of this paper can be added to that list. 

To date, only the branching ratios of the decays $\bd\to\kp\km$ and $\bs\to\pip\pim$ have been measured. If/when the direct and indirect CP asymmetries of these decays are measured, this will give us additional information that may help shed light on the U-spin puzzle.

\bigskip
\noindent
{\bf Acknowledgments}: This work was financially supported by the National Science Foundation, Grant No.\ PHY-2013984 (BB) and by NSERC of Canada (SK, DL, NP).

\end{document}